\documentclass[12pt]{iopart}

\expandafter\let\csname equation*\endcsname\relax
\expandafter\let\csname endequation*\endcsname\relax

\normalsize

\usepackage[utf8]{inputenc}
\usepackage[british]{babel}
\usepackage{braket}
\usepackage[colorlinks=true, linkcolor=blue, citecolor=blue, urlcolor=blue]{hyperref}
\usepackage{booktabs}
\usepackage{amsthm}
\usepackage{amsfonts}
\usepackage{amsopn}
\usepackage{amsmath}
\usepackage{bm}
\usepackage{amssymb}
\usepackage[pdftex]{graphicx}
\usepackage[Export]{adjustbox}
\usepackage{wrapfig}
\usepackage{float}
\usepackage{xfrac}
\usepackage{microtype}
\usepackage[dvipsnames]{xcolor}
\usepackage[squaren]{SIunits}
\usepackage{footnotebackref}
\usepackage{subfigure}
\usepackage{txfonts}
\usepackage{bbold}
\usepackage[section]{placeins}
\usepackage{ulem}
\usepackage{caption}

\newcommand{\hyhref}[2]{\hyperref[#1]{#2}}

\begin{document}

\title{Driving Enhanced Exciton Transfer by Automatic Differentiation}

\author{E. Ballarin}
\address{Department of Mathematics, Informatics, and Geoscience, University of Trieste, via Alfonso Valerio 2, 34127 Trieste, Italy}
\author{D. A. Chisholm}
\address{Centre for Quantum Materials and Technologies, School of Mathematics and Physics, Queen's University Belfast, BT7 1NN, United Kingdom}
\author{A. Smirne}
\address{Dipartimento di Fisica `Aldo Pontremoli', Universit\`a degli Studi di Milano, via Celoria 16, 20133 Milan, Italy}
\address{Istituto Nazionale di Fisica Nucleare, Sezione di Milano, via Celoria 16, 20133 Milan, Italy}
\author{M. Paternostro$^{\star}$}
\address{Universit\`a degli Studi di Palermo, Dipartimento di Fisica e Chimica - Emilio Segr\`e, via Archirafi 36, 90123 Palermo, Italy}
\address{Centre for Quantum Materials and Technologies, School of Mathematics and Physics, Queen's University Belfast, BT7 1NN, United Kingdom}
\author{F. Anselmi$^{\star}$}
\address{Department of Mathematics, Informatics, and Geoscience, University of Trieste, via Alfonso Valerio 2, 34127 Trieste, Italy}
\address{MIT, 77 Massachusetts Ave, Cambridge, 02139, MA, USA}
\ead{fabio.anselmi@units.it}
\author{S. Donadi$^{\star}$}
\address{Centre for Quantum Materials and Technologies, School of Mathematics and Physics, Queen's University Belfast, BT7 1NN, United Kingdom}
\address{Istituto Nazionale di Fisica Nucleare, Sezione di Trieste, via Alfonso Valerio 2, 34127 Trieste, Italy}
\ead{s.donadi@qub.ac.uk}

\pagebreak

\begin{abstract}
We model and study the processes of excitation, absorption, and transfer in various networks. The model consists of a harmonic oscillator representing a single-mode radiation field, a qubit acting as an antenna, a network through which the excitation propagates, and a qubit at the end serving as a sink. We investigate how off-resonant excitations can be optimally absorbed and transmitted through the network. Three strategies are considered: optimising network energies, adjusting the couplings between the radiation field, the antenna, and the network, or introducing and optimising driving fields at the start and end of the network. These strategies are tested on three different types of network with increasing complexity: nearest-neighbour and star configurations, and one associated with the Fenna-Matthews-Olson complex. The results show that, among the various strategies, the introduction of driving fields is the most effective, leading to a significant increase 
in the probability of reaching the sink in a given time. This result remains stable across networks of varying dimensionalities and types, and the driving process requires only a few parameters to be effective.   
\end{abstract}

\maketitle

\section{Introduction}
Complex systems are composed of several interacting components with a behaviour that is not immediately predictable from the characteristics of its parts, and often exhibits emergent phenomena. They include, among others, social~\cite{boguna2004models, borgatti2009network} and economic structures~\cite{mantegna1995scaling, bonanno2004networks} as well as biological complexes~\cite{davidson2005gene, jeong2001lethality}, and are currently the subject of intense study: either to understand the emergence of new phenomena or as tools to analyse real-world complex scenarios.
Complex systems are nowadays receiving much attention and study also in the quantum context~\cite{nokkala2024complex}. In fact, it is now possible to engineer and manipulate multipartite quantum systems in many experimental settings, reaching sizes where their complexity becomes significant and practically relevant~\cite{van2016optimal, lodahl2017quantum}. Complexity has been thus identified as a precious resource for several quantum tasks, ranging from communication to computation and metrological ones~\cite{apers2022quadratic}.

One paradigmatic example of complex quantum dynamics is that of continuous-time quantum walks (CTQWs)~\cite{mulken2011continuous}, where a single quantum walker flows throughout a physical system, represented by a complex network.
CTQWs have been used to model the excitation transfer in bio-molecule complexes~\cite{caruso2009highly, chin2010noise, chisholm2021stochastic}, offering insight into the intricate interplay between environmental effects, the network structure, and surviving quantum features in determining the performance of the transfer process.
In particular, because of the dissipative role of the environment, fast excitation transfer is needed to maximise its efficiency.

In this paper, we investigate excitation transfer in complex networks with the purpose of optimising transfer speed. Inspired by the excitation transfer process in a photovoltaic system, our goal is to describe, using a simple model, the absorption of a photon, its conversion into an excitation, and subsequent transfer along a network.
We focus in particular on the case where two of the system components are subject to an external driving, and make use of algorithmic differentiation and gradient-based optimisation techniques to find the optimal driving.

We find that such a relatively simple degree of control can greatly enhance the efficiency of the process. We also find that such a novel approach can be coupled with previously studied methods~\cite{davidson2021principles, sgroi2024efficient} to yield better results compared to already optimised protocols.
Furthermore, the inclusion of the absorption mechanism (a qubit acting as an antenna) allows us to have a more general picture compared to the scenario typically considered in the literature \cite{davidson2021principles}.

The engineering of protocols to enhance the transport efficiency in quantum networks can be challenging due to the complex nature of the system. Moreover, while other control strategies, such as changing the environmental dissipation or engineering the network couplings, have been proven to be effective, their physical implementation would be far from trivial or even unfeasible.
On the other hand, enacting a local control only on two of the system nodes would prove much more feasible at the experimental level. Additionally, as we are implementing control on a fixed number of qubits, the number of parameters to optimise is constant regardless of the size of the network, while the complexity of alternative optimisation strategies may scale unfavourably with the network size.

This establishes a viable pathway to optimisation with the feature of being more computationally agile than other strategies used in the past~\cite{Munro,Zwick_2014,Ai_2014,Xiang_2013}, making it a valuable tool for future optimisation procedures in similar settings.

\section{General settings and methods}

We aim to describe the photovoltaic process through a minimal model, which involves some radiation incident on an antenna that absorbs the incoming energy and converts it into an excitation transmitted across a network. The $N^\text{th}$ site of the latter is assumed to be connected to a \textit{sink-like} system, where the excitation is stored. 

\begin{figure}[h!]
    \centering
    \includegraphics[width=10cm]{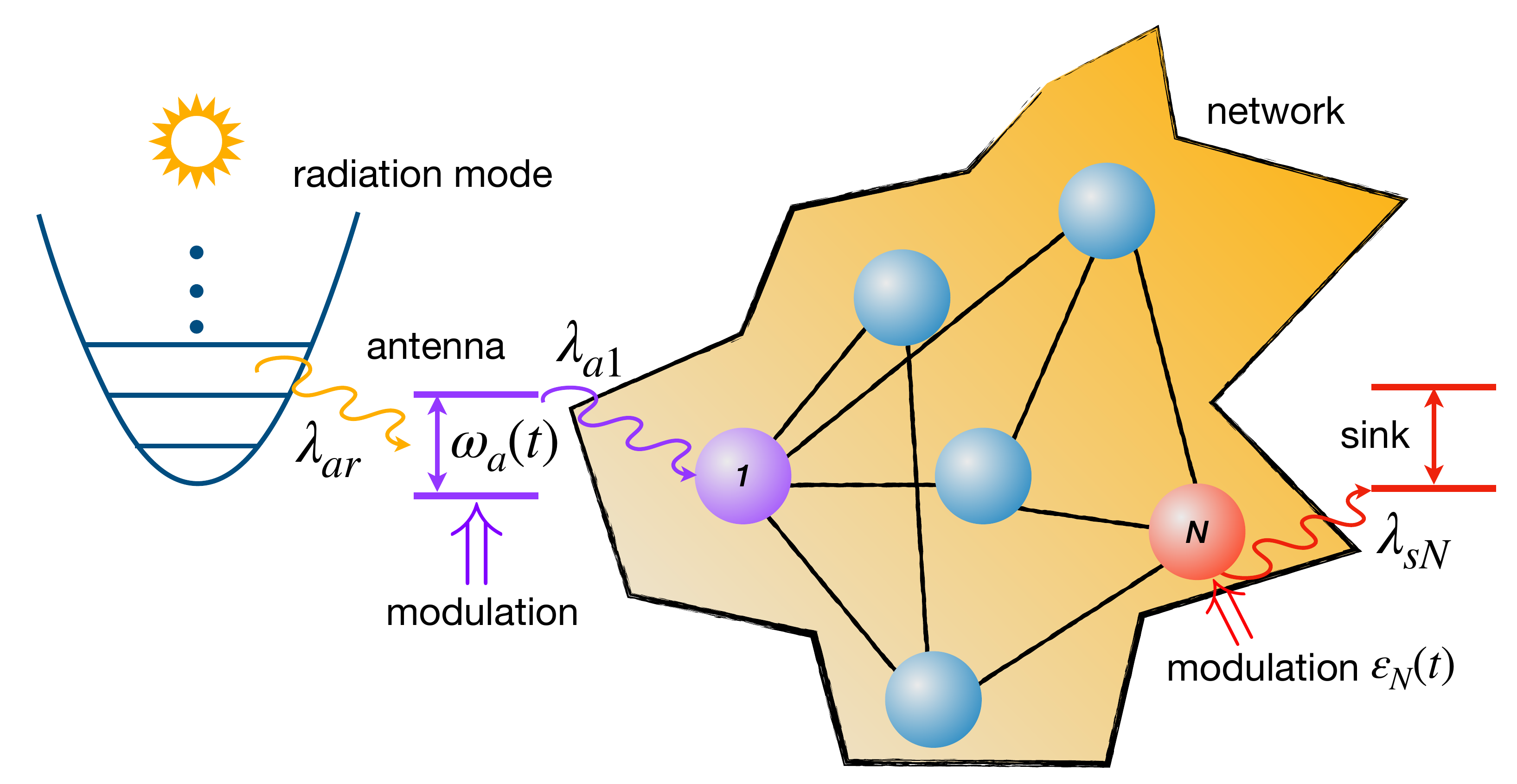}
    \caption{\small Schematic of the model. One mode of the radiation interacts with the antenna (qubit 1, subscript `a'), exciting its ground state. The excitation then jumps to the first site of the network and travels through it, up to the last site ($N$), which is connected to a sink (qubit 2, subscript `s').}
    \label{fig1}
\end{figure}

In modelling such elements, we must strike a delicate balance between accuracy and simplicity to ensure the possibility of grasping an intuition of the physics underpinning the overall process. 
A sketch of such arrangements, where both the antenna and the sink are modelled as
two-level systems, is shown in \autoref{fig1}. The incoming radiation is considered as a single mode, represented by a harmonic oscillator initially in its first excited state. This oscillator interacts through a hopping interaction with qubit 1, which acts as an antenna, and excites it. Qubit 1 is then connected to the first site of a network with $N$ sites. For the network, we choose to work in the subspace of a single excitation, which is sufficient for our study and significantly reduces the dimension of the Hilbert space (which is $N+1$, compared to $2^N$, for a network with $N$ sites represented by qubits). We also include non-unitary terms in the network to account for possible local dephasing. Finally, the last site of the network is connected through non-unitary dynamics to qubit 2, which acts as the sink. The Bohr frequency of qubit 1 and the energy of the last site of the network will generally be time-dependent. The reason for maintaining such a time dependence is that, as we will discuss in detail, one possible way to increase energy transfer along the network is to introduce suitable time-dependent drivings.

Accordingly, the model is described by a Lindblad Master Equation \cite{lindblad1976generators, gorini1976completely} of the form: 
\begin{equation}\label{me}
\frac{d\hat{\rho}(t)}{dt}=-\frac{i}{\hbar}\left[\hat{H},\hat{\rho}(t)\right]+L_{N}[\hat{\rho}(t)] + L_{S}[\hat{\rho}(t)],
\end{equation}
where the Hamiltonian is a sum of several contributions:
\begin{equation}\label{H}
\hat{H}=\hat{H}_{R}+\hat{H}_{A}(t)+\hat{H}_{N}(t)+\hat{H}_{AR}+\hat{H}_{A1}.
\end{equation}
The Hamiltonian of the single radiation mode with frequency $\omega_r$ is $\hat{H}_{R}=\hbar\omega_{r}\hat{a}^{\dagger}\hat{a}$. The Hamiltonian of the antenna is that of a simple two-level system, i.e.
\begin{equation}
\hat{H}_{A}(t)=\hbar\omega_{a}(t)\left(|e_a\rangle\langle e_a|-|g_a\rangle\langle g_a|\right),
\end{equation}
where $\ket{e_a}=\begin{pmatrix}1\\0\end{pmatrix}$ and $\ket{g_a}=\begin{pmatrix}0\\1\end{pmatrix}$ are the excited and ground state.
We take the frequency $\omega_a(t)$ to be the time-dependent function 
\begin{equation}\label{wq}
\omega_{a}(t)=\omega_{a}+\sum_{i=1}^R A_{i}\sin(\nu_{i}\,t+\phi_i)
\end{equation}
with $\omega_a$ the Bohr frequency of the $\ket{e_a}\leftrightarrow\ket{g_a}$ transition in the absence of any modulation, and $A_i, \nu_i,\phi_{i}$ real parameters that enter the harmonic-based decomposition of the modulation strategy that we will aim at optimising. 

The Hamiltonian of the network is given by
\begin{equation}\label{HN}
\hat{H}_{N}=\sum_{i=1}^{N-1}\varepsilon_{i}|i\rangle\langle i|+\varepsilon_{N}(t)|N\rangle\langle N|+\sum_{i\neq j=1}^{N}V_{ij}\left(|i\rangle\langle j|+|j\rangle\langle i|\right),
\end{equation}
where $\varepsilon_i$ are the energies of the sites and $V_{ij}$ the couplings of the interaction between sites $i$ and $j$. The last site $N$, which is connected to the sink, has an energy which can be time-dependent to allow the introduction of a driving, which similarly to Eq. \eqref{wq} takes the form 
\begin{equation}\label{epsN}
\varepsilon_{N}(t)=\varepsilon_{N}+\sum_{i=1}^R B_{i}\sin(\mu_{i}\,t+\theta_i),
\end{equation}
with $\varepsilon_N$, $B_i$, $\mu_i$, and $\theta_i$ real parameters. Note that the Hilbert space of the network, in addition to the $N$ states representing its sites, contains also a vacuum state $|0\rangle$ with energy $\varepsilon_0=0$ that does not interact with any other site of the network. This site essentially represents the absence of excitations in the network, and it only interacts with qubit 1. The reason for its introduction is that, contrary to what is typically done where one assumes that the excitation is in the first site of the network at time $t=0$, here we are also modelling the absorption of the excitation. Hence, we need a state representing the absence of excitation in the network, which we model by adding an additional site that does not interact with the others.

The radiation-antenna and antenna-network interactions are described by
\begin{equation}\label{HqR}
\hat{H}_{AR}=\lambda_{ar}|e_a\rangle\langle g_a|\otimes \hat{a}+ h.c.
~~~~~~\text{and}~~~~~~
\hat{H}_{A1}=\lambda_{a1}|e_a,0\rangle\langle g_a,1|+h.c.,
\end{equation}
respectively. Such hopping Hamiltonians 
rule the coherent transfer of excitations between the radiation mode and the antenna (at rate $\lambda_{ar}$), and between the antenna and the first site of the network (at rate $\lambda_{a1}$). 

Going back to Eq.~\eqref{me}, we introduce the incoherent term
\begin{equation}
L_{S}[\hat{\rho}]=\lambda_{sN}\left(P_{g_s,N}|e_{s},0\rangle\langle e_{s},0|-\frac{1}{2}\{|g_{s},N\rangle\langle g_{s},N|,\hat{\rho}\}\right)\text{,}
\end{equation}
with $P_{g_s,N}=\langle g_{s},N|\hat{\rho}|g_{s},N\rangle$ the population of state $\ket{g_s,N}$ of the sink-site $N$ compound. This describes the one-way mechanism through which an excitation populating site $N$ is transferred to the sink at a rate $\lambda_{sN}$. We also have the local dephasing mechanism 
\begin{equation}
L_{N}[\hat{\rho}]=\lambda_N \left( \sum_{j=1}^{N}|j\rangle\langle j|\hat{\rho}|j\rangle\langle j|-\hat{\rho}\right)
\end{equation}
with $|j\rangle$ representing the state where site $j$ of the network is populated, and $\lambda_N$ the dephasing rate, assumed for simplicity to be the same for all sites.

In the next section, we will study several ways to improve excitation transfer from the radiation to the sink. In all the case-studies, we will take as initial state
\begin{equation}\label{in_state}
\hat{\rho}(0)=|1_r\rangle\langle1_r|\otimes|g_a\rangle\langle g_a|\otimes|0\rangle\langle0|\otimes|g_s\rangle\langle g_s|,
\end{equation}
i.e. the oscillator is in its first excited state and all other elements of the model are in their ground state.

We evolve this state, using a fourth-order Runge-Kutta method (see \cite{bookRK}, p. 215), according to Eq. \eqref{me} for different networks. Then, we optimise different sets of parameters in order to maximise the probability that the excitation reaches the sink in the shortest time.

\section{Analysis and results}

The model introduced in the previous section depends on many potentially tunable parameters. In what follows, we will optimise subsets of these parameters (keeping the remaining fixed), to maximise the time-integrated probability that an excitation reaches the sink within a given time.

We focus on three choices for the parameters to be optimised.
In the first case, we optimise the drivings $\omega_a(t)$ and $\varepsilon_N(t)$; in the second case, the couplings $\lambda_{ar}$ and $\lambda_{a1}$; in the last case, the energies of the network $\varepsilon_i$.
In each of those cases, we compare the setting in which selected parameters are optimised, with the corresponding unoptimised scenario.

Parameter optimisation is carried out using gradient-based methods, relying on automatic differentiation \cite{automaticdiff} provided by the Python library \textit{PyTorch} \cite{PaszkePytorch} and using the \textit{Adam} optimiser \cite{KingBa15}. Specifically, the learning rate of the optimiser has been chosen in a case-dependent fashion, according to the empirical complexity of the loss landscape \cite{Shallue2023HowTo}.

In detail, the parameters are optimised in such a way to maximise the time-integrated probability for the excitation to reach the sink, i.e.
\begin{equation}\label{Ip} I_P(T_L):=\int_0^{T_L}p_{sink}(t)\;dt\text{,}
\end{equation}
where $p_{sink}(t)=\textrm{Tr}[|e_s\rangle\langle e_s|\hat{\rho}(t)]$
with $\hat{\rho}(t)$ the total density matrix at time $t$, $|e_s\rangle\langle e_s|$ the projector on the excited state of the sink and $T_L$ the time of evolution used as part of the optimisation of the parameters. The rationale for such a choice is that by maximising $I_P(T_L)$ instead of the final probability $p_{sink}(T_L)$, we can identify a set of parameters that not only maximises $p_{sink}(T_L)$, but also favours solutions where $p_{sink}(t)$ increases significantly at earlier times, i.e., those where the excitation reaches the sink more quickly.

We perform our analyses on three different types of network: one describing a nearest neighbour (NN) interaction, a star network (SN) where one of the sites is connected to all the others, and one modelling the Fenna–Matthews–Olson (FMO) complex \cite{engel2007evidence, cheng2009dynamics}.


\subsection{Nearest neighbour network}

We start by considering the case of a NN network, which has the Hamiltonian:
\begin{equation}
    H_{N}=\left(\begin{array}{ccccccc}
0 & 0 & 0 & ... & 0 & 0 & 0\\
0 & 0.5 & 1 & ... & 0 & 0 & 0\\
0 & 1 & 0.5 & ... & 0 & 0 & 0\\
... & ... & ... & ... & ... & ... & ...\\
0 & 0 & 0 & ... & 1 & 0.5 & 1\\
0 & 0 & 0 & ... & 0 & 1 & 0.5
\end{array}\right).
\end{equation}
The remaining parameters used in the simulation are summarised in \autoref{t1}. We set $\hbar=1$, and therefore the parameters with the dimension of an energy are expressed in units of $\omega_a$; times are expressed in units of $\omega^{-1}_a$. We choose as the initial state that specified in \eqref{in_state}.
\begin{table}[t]
\centerline{
      \begin{tabular}{@{}c|c|c|c|c|c|c|c@{}}
        \toprule
        $\omega_a T_L$ & $\omega_a T$  & $\frac{\omega_r}{\omega_a}$      & $N$      & $\frac{\lambda_{ar}}{\omega_a}$ & $\frac{\lambda_{a1}}{\omega_a}$ & $\frac{\lambda_{N}}{\omega_a}$ & $\frac{\lambda_{sN}}{\omega_a}$ \\ \midrule
        30    & 1200 & \{0.264, 15\}           & \{4, 8\} & 1              & 1              & 0.1             & 0.05        \\
        \bottomrule
      \end{tabular}
  }
\caption{Values of model parameters used to run the simulations; times are in units of $\omega_a^{-1}$ and frequencies in units of $\omega_a$. $T_L$ is the evolution time used within the optimisation process; after optimisation, we test the performance of the learned system by evolving it further up to time $T$; $\omega_r$ is the frequency of the radiation, $\omega_a$ the Bohr frequency of qubit 1 (when there is no driving), $N$ the number of sites in the network, $\lambda_{ar}$ and $\lambda_{a1}$ the values of the couplings (when not optimised), $\lambda_N$ the value of the dephasing constant and $\lambda_{sN}$ the value of the sink constant.}
  \label{t1}
\end{table}

All simulations are run for a total time of $T=1200$, which is significantly longer than the evolution used within the optimisation process ($T_L=30$). In fact, we verified that a further increase in $T_L$ does not result in any significant improvement in the integral of sink probability. The coupling $\lambda_S$ is taken to be an order of magnitude smaller than all other couplings to limit the resulting Zeno effect, which would significantly influence the flow of excitation along the network.

We start by considering a network with $N=4$ sites and an incident radiation with a frequency of
$\omega_r=0.264$. This baseline value is chosen as it maximises the value of $I_P(T_L)$ in Eq. \eqref{Ip} for the parameters shown in \autoref{t1}.

\begin{figure*}[htbp]
    \includegraphics[width=0.31\textwidth]{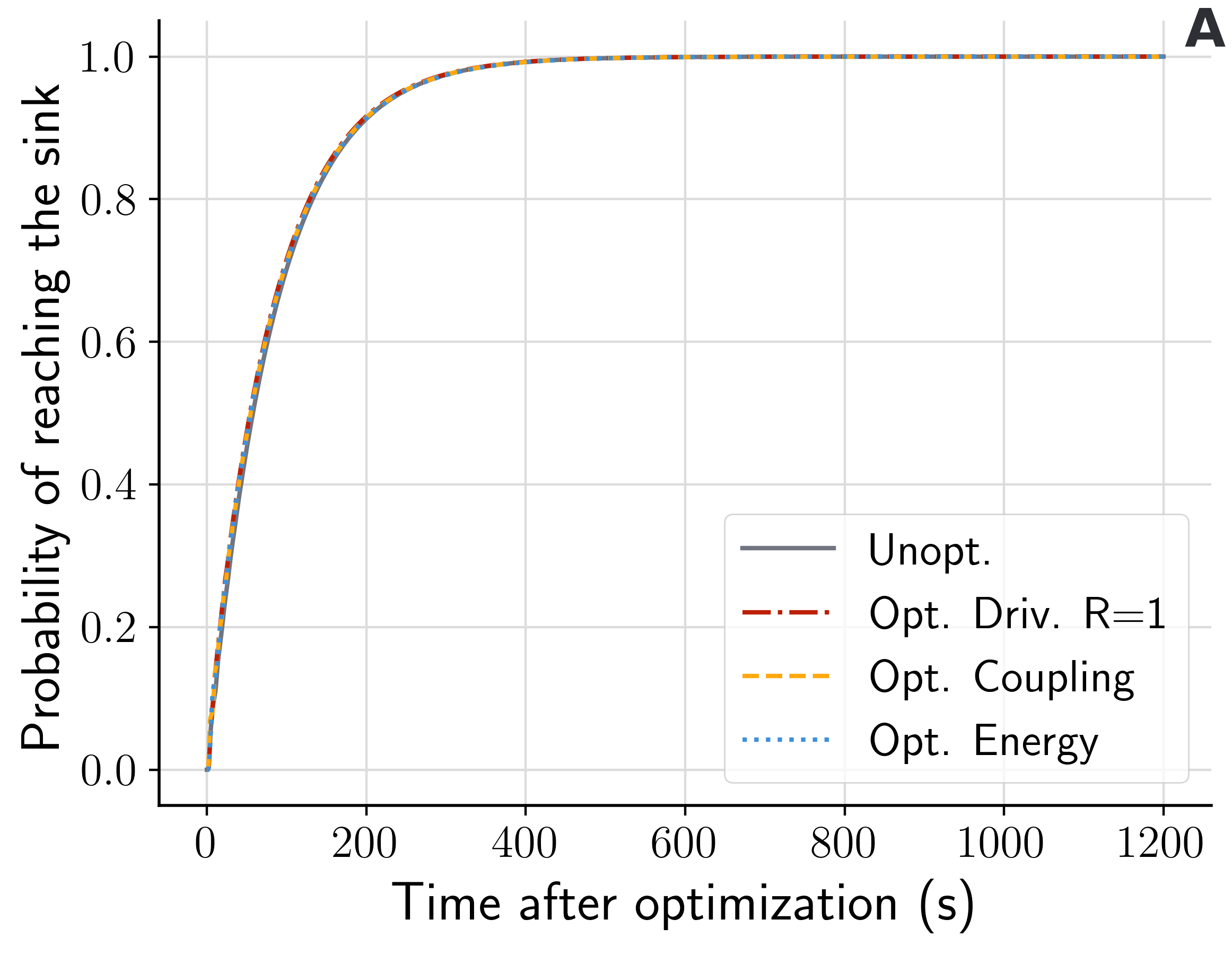}
        \includegraphics[width=0.31\textwidth]{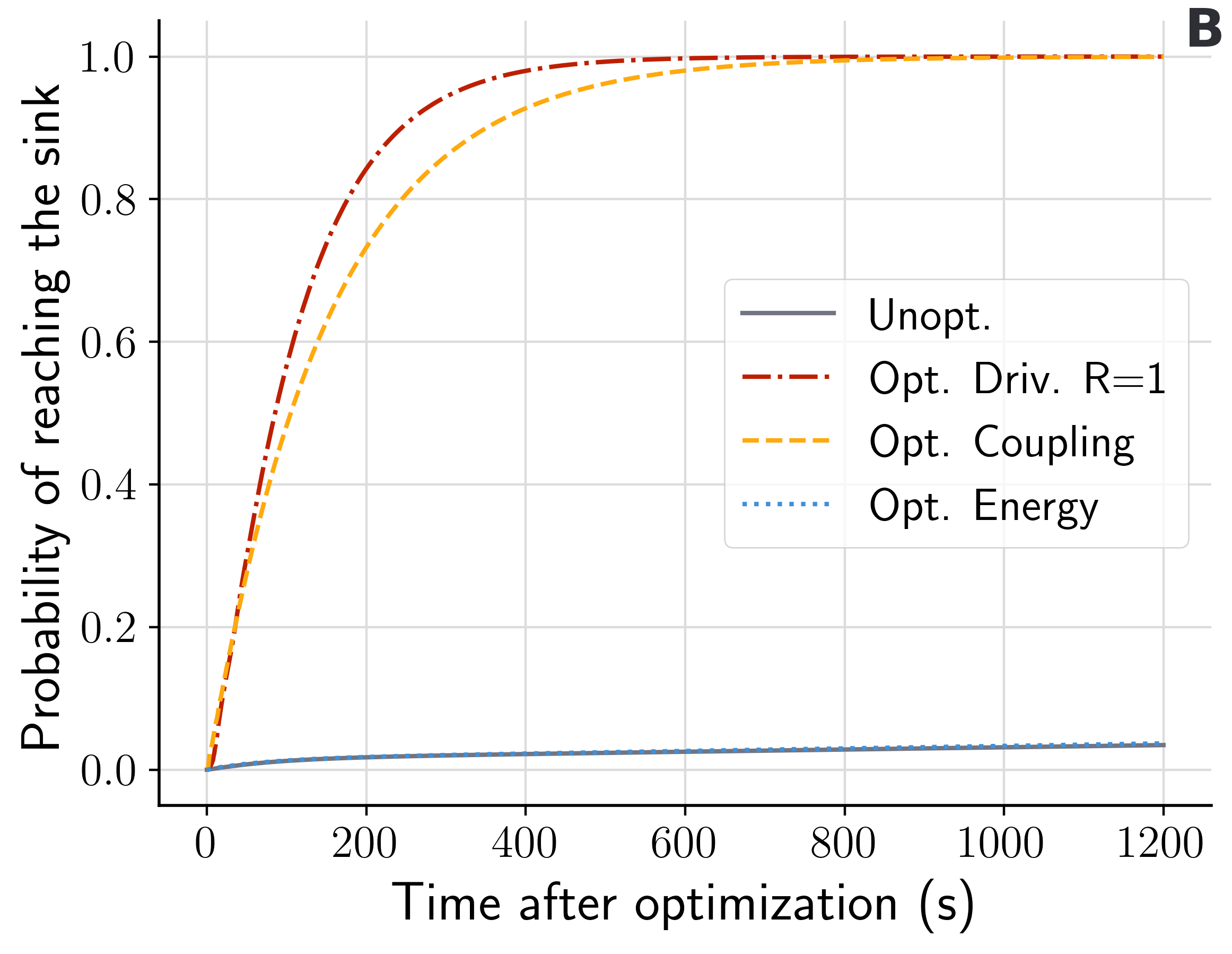}
           \includegraphics[width=0.31\textwidth]{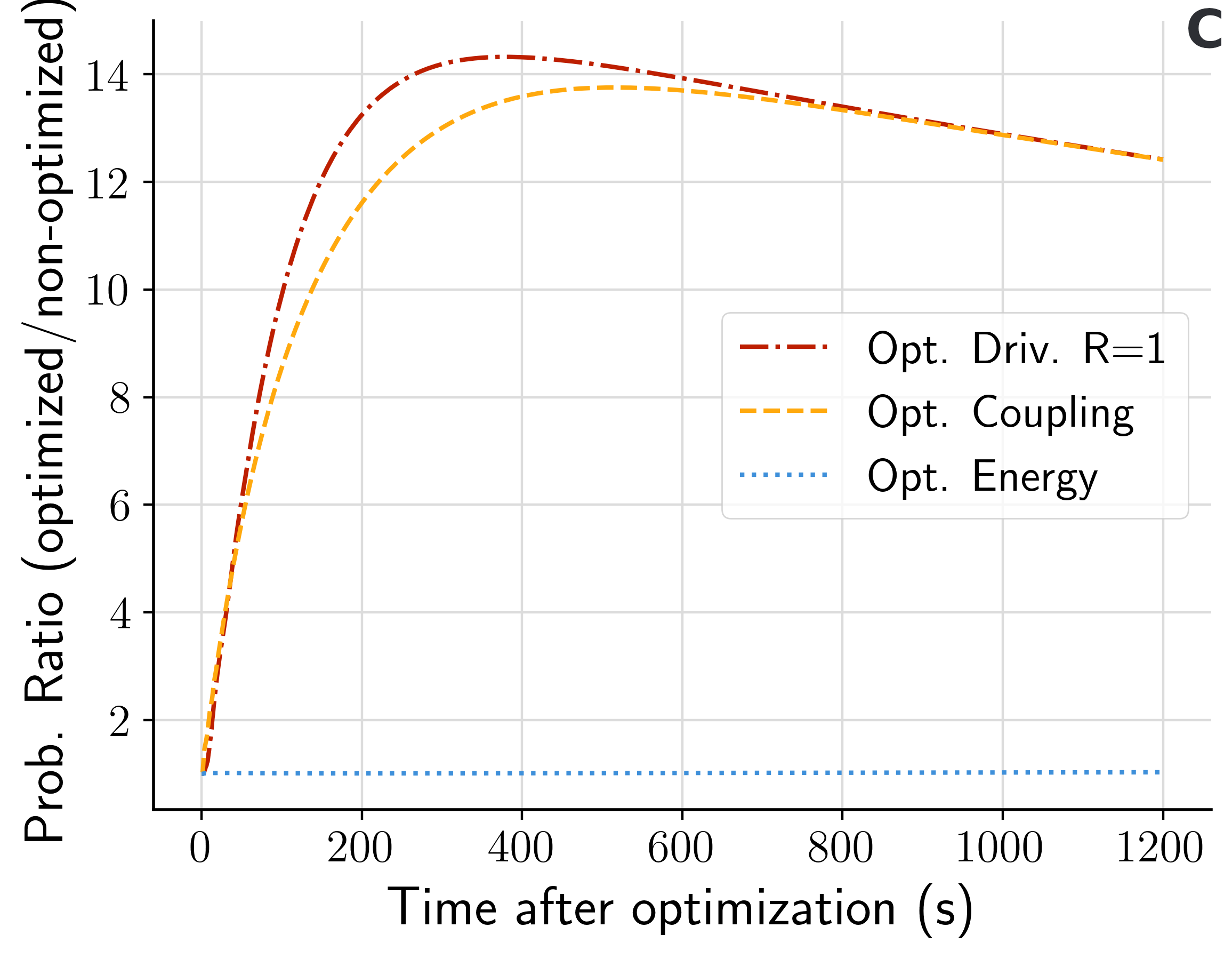}
    \caption{\small Comparison among unoptimised network, optimised driving (with $R=1$ in Eqs.~\eqref{wq} and \eqref{epsN}), optimised coupling, and optimised energies for a NN network of $N=4$ sites. (A) Probability of reaching the sink when the frequency of the mode is $\omega_r=0.264$ (resonant case): the improvement with respect to the unoptimised network is minimal with all the three methods. (B)  Probability of reaching the sink when the frequency of the mode is $\omega_r=15$ (off-resonant case): the improvement is relevant when optimising the couplings or the drivings. (C) Ratio between the probability of reaching the sink with the three optimisation strategies and the probability of reaching the sink without optimisation.}
    \label{F02}
    
    \end{figure*}

\autoref{F02}(A) shows the probability of the excitation reaching the sink as a function of time, comparing the case without any optimisation (grey curve) to the cases where the driving parameters (red curve),  the couplings (yellow curve) and the network energies (blue curve) are optimised. We see that none of the three optimisation strategies is particularly effective compared to the baseline, because $\omega_r=0.264$ is already optimal. One might argue that, by considering more complex drivings where more terms ($R>1$) are accounted for in the series in Eqs. (\ref{wq}) and (\ref{epsN}) the performance could potentially improve. 
We found out that this is not the case and that, interestingly, increasing the number of terms in the drivings up to $R=7$ (corresponding to a total of 21 real parameters to learn for each driving term) the network after optimisation performs worse than the original. This behaviour is probably due to the significant increase of the complexity of the solutions landscape and therefore to a more difficult optimisation problem.

We now turn to a more interesting case by selecting a different frequency for the incident radiation, $\omega_r = 15$. This particular value is chosen to be significantly different from the baseline at $\omega_r = 0.264$, at which the network is known to perform well. By using an incident photon with significantly higher energy, we expect it to be far off-resonance, making the optimisation of tunable parameters, or the introduction of external driving, way more effective. \autoref{F02}(B) shows that this is the case. In fact, the probability of reaching the sink is increased by more than an order of magnitude when the drivings or the couplings are learnt (see \autoref{F02}(C)). On the other hand, optimising the network energies is not so effective, as shown by the dotted blue line in \autoref{F02}(C). 

To further test the strategy based on introducing and optimising drivings, we studied how it changes by increasing the number of terms in the drivings $R=1,2,7$ or by considering network of different sizes $N=4,6,8$. As one can see in \autoref{Fapp}(A) of Appendix 1, no sensible improvement is reached by increasing $R$. Also, as shown in \autoref{Fapp}(B) of Appendix 1, the conclusions we draw for the case $N=4$ are qualitatively confirmed for larger networks, showing that driving optimisation is effective over larger networks.

\begin{figure*}[htbp]
    \centering
    \subfigure{
        \includegraphics[width=0.45\textwidth]{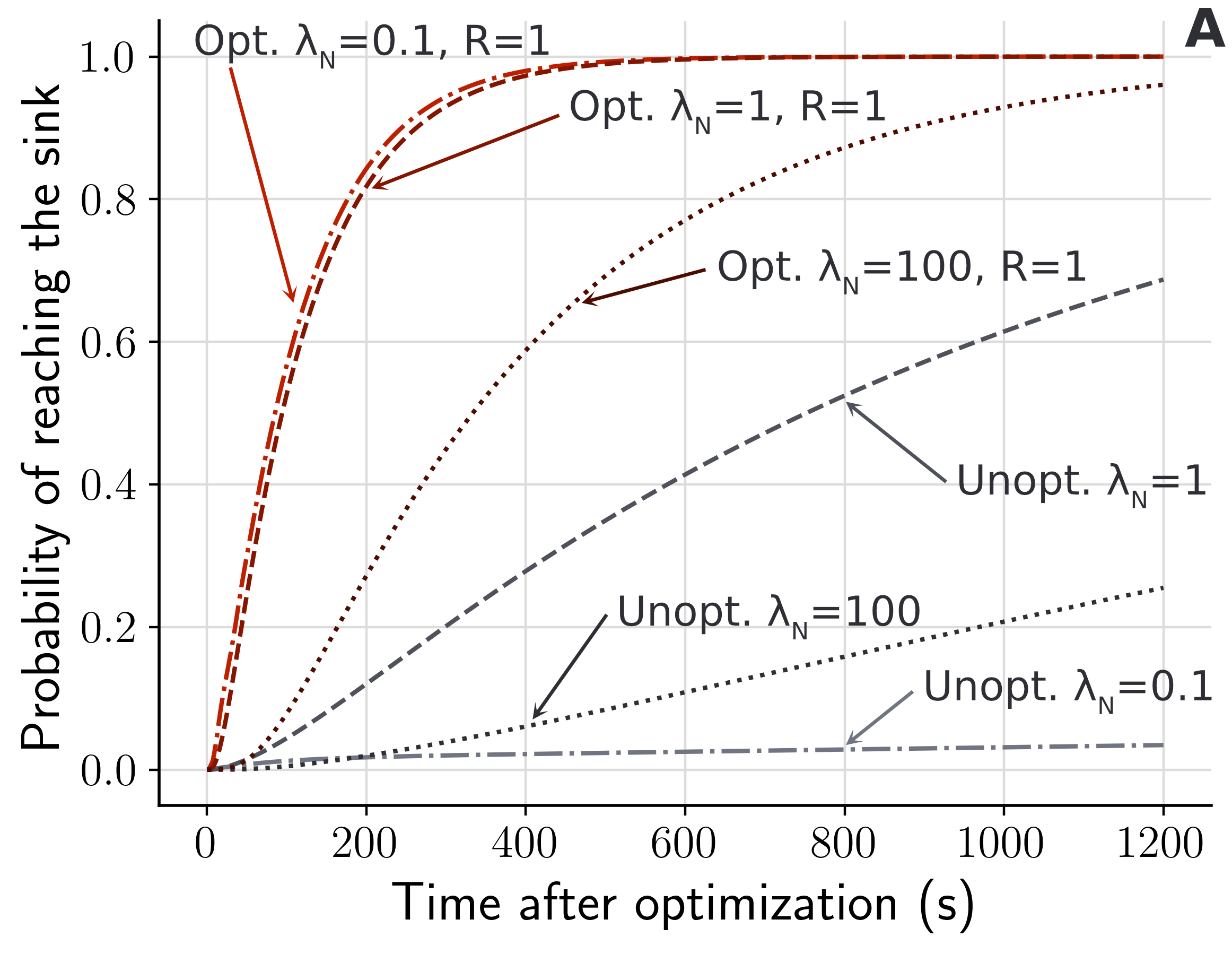}
    }
    \subfigure{
        \includegraphics[width=0.45\textwidth]{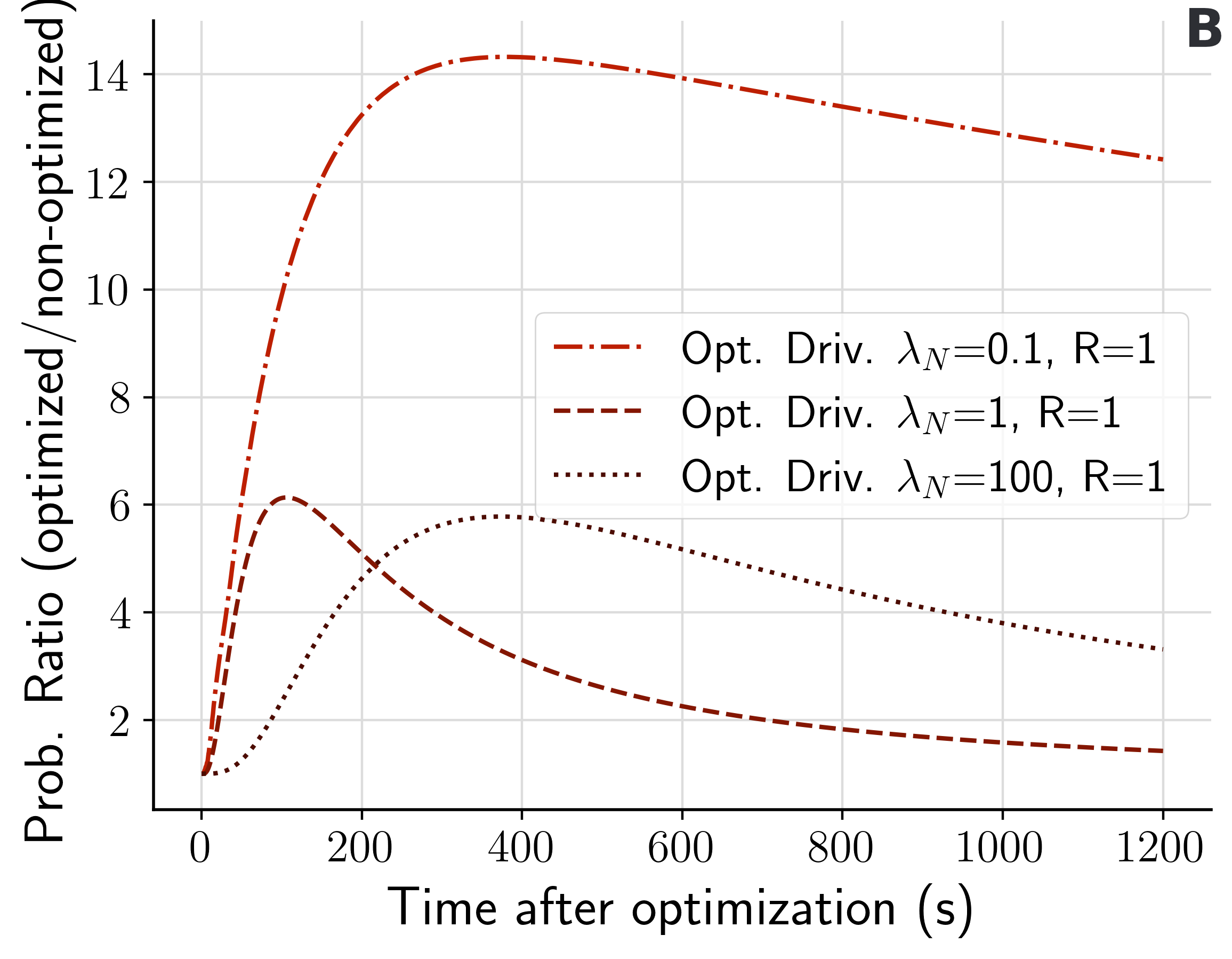}
    }
    \caption{(A) Probability of reaching the sink for unoptimised network (Unopt) and one where the drivings where optimised (Opt) ($\omega_r=15$ and $R=1$ in Eqs.~\eqref{wq} and \eqref{epsN}). The simulation were run for different values of the dephasing parameter $\lambda_N= 0.1,1,100$. (B) Ratio between the probability of reaching the sink with the driving optimisation and the probability of reaching the sink without optimisation for the three values of $\lambda_N$.}
    \label{F10}
\end{figure*}

Moreover, we tested all three methods under increasing levels of noise. In \autoref{Fapp}(C) of Appendix 1, we repeated the same analysis as in \autoref{F02}(B), but with $\lambda_N = 1$ (instead of $0.1$), finding no significant differences. We then focused on optimising only the driving, for increasing levels of noise ($\lambda_N = 0.1, 1, 100$). The results presented in \autoref{F10} demonstrate that driving optimisation remains highly effective. They also reveal an interesting phenomenon: increasing noise levels in an unoptimised system may lead to a higher integrated sink probability. This appears to be an instance of noise-assisted transport \cite{plenio2008dephasing}.

\subsection{Star Network}
In this section we consider a different type of network, called \textit{star network} (SN), where one of the sites (the second, in our case) is connected to all the others. The goal of such an analysis is to check whether the conclusions reached in the previous scenario also extend to other network setups.

The Hamiltonian of the star network is: 
\begin{equation}\label{HNstar}
    H_{star}=\left(\begin{array}{cccccccc}
0 & 0 & 0 & 0 & 0 & 0 & 0 & 0\\
0 & 0.5 & 1 & 0 & 0 & 0 & 0 & 0\\
0 & 1 & 0.5 & 1 & 1 & 1 & 1 & 1\\
0 & 0 & 1 & 0.5 & 0 & 0 & 0 & 0\\
0 & 0 & 1 & 0 & 0.5 & 0 & 0 & 0\\
0 & 0 & 1 & 0 & 0 & 0.5 & 0 & 0\\
0 & 0 & 1 & 0 & 0 & 0 & 0.5 & 0\\
0 & 0 & 1 & 0 & 0 & 0 & 0 & 0.5
\end{array}\right).
\end{equation}

\begin{figure*}[htbp]
    \includegraphics[width=0.31\textwidth]{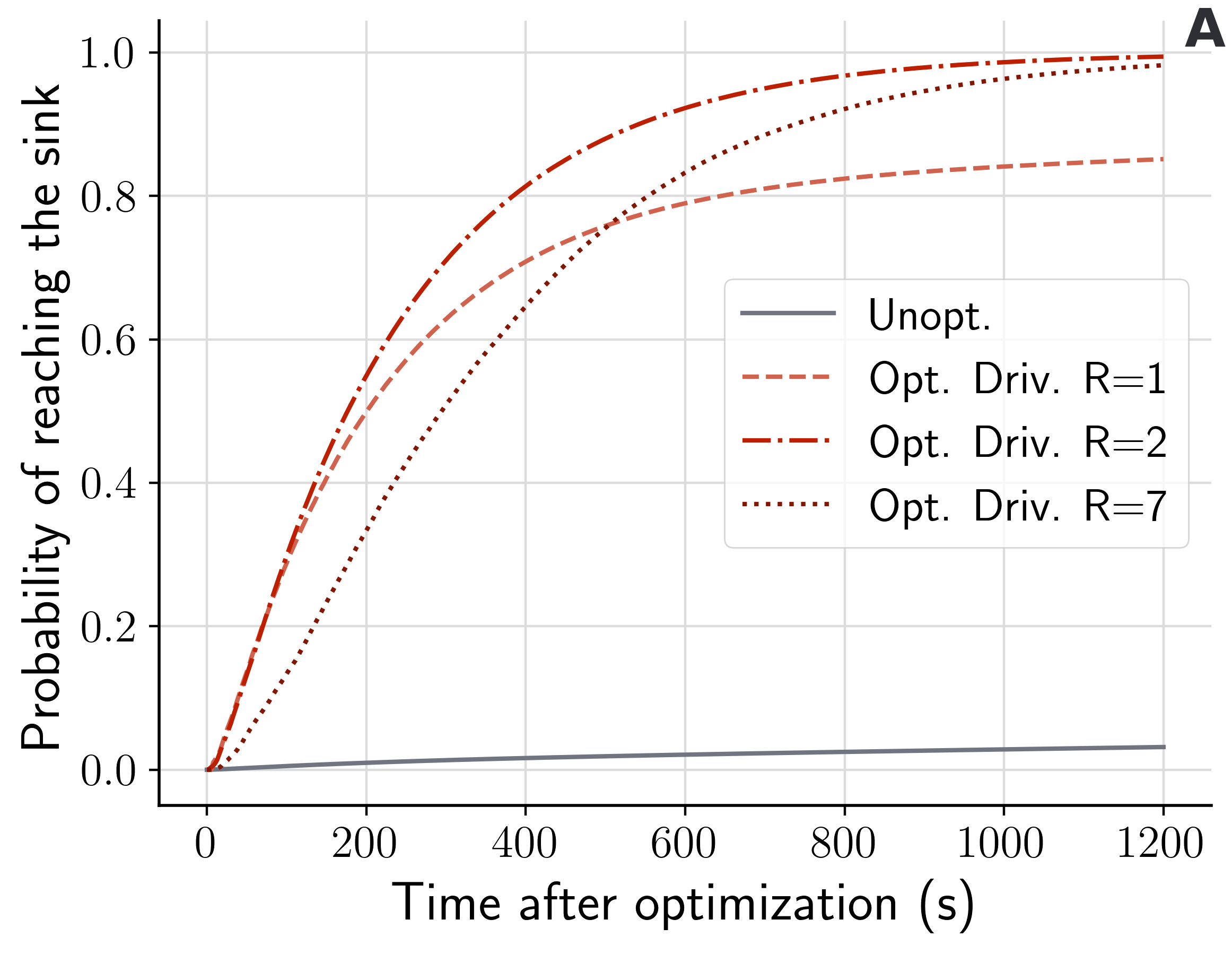}
        \includegraphics[width=0.31\textwidth]{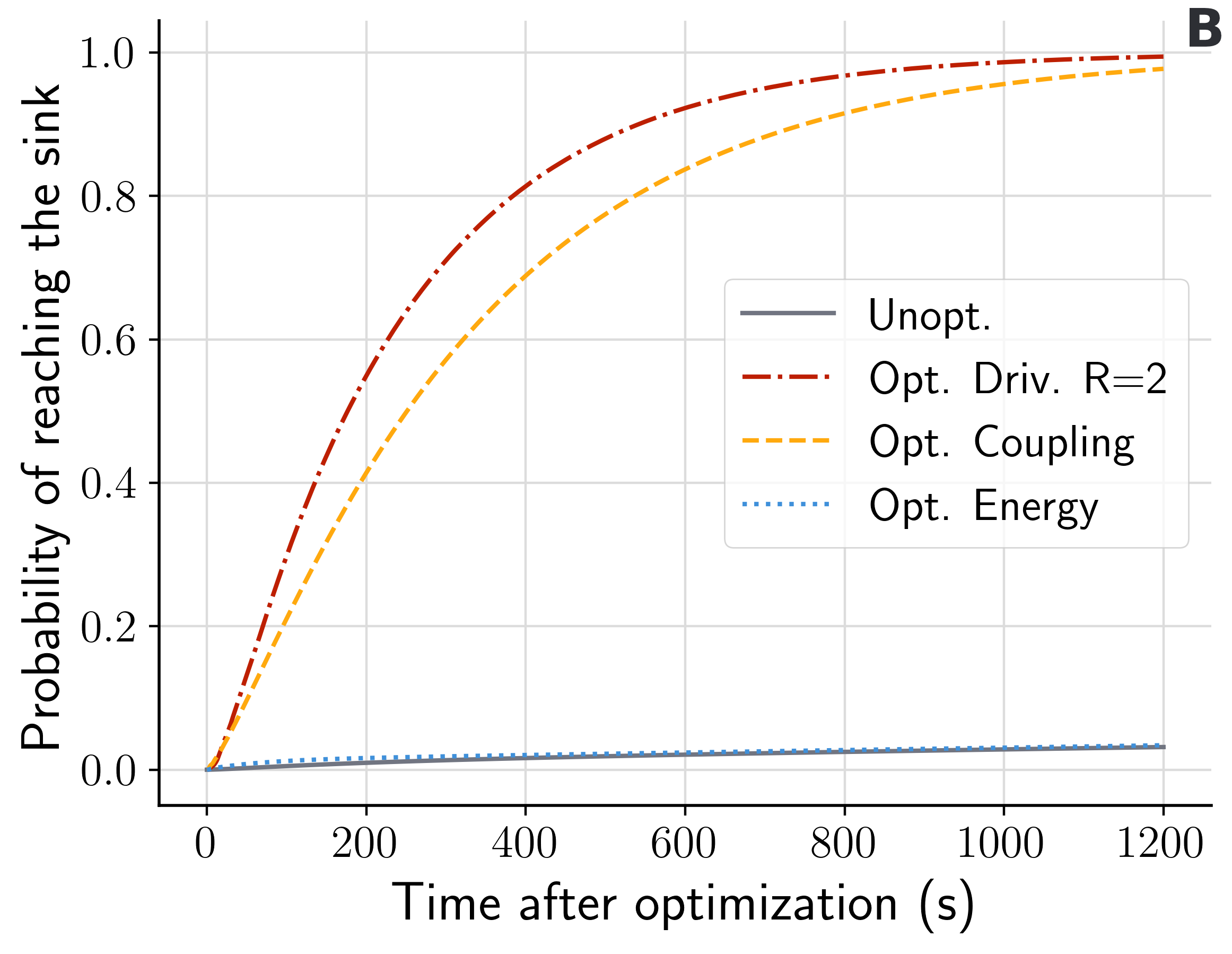}
           \includegraphics[width=0.31\textwidth]{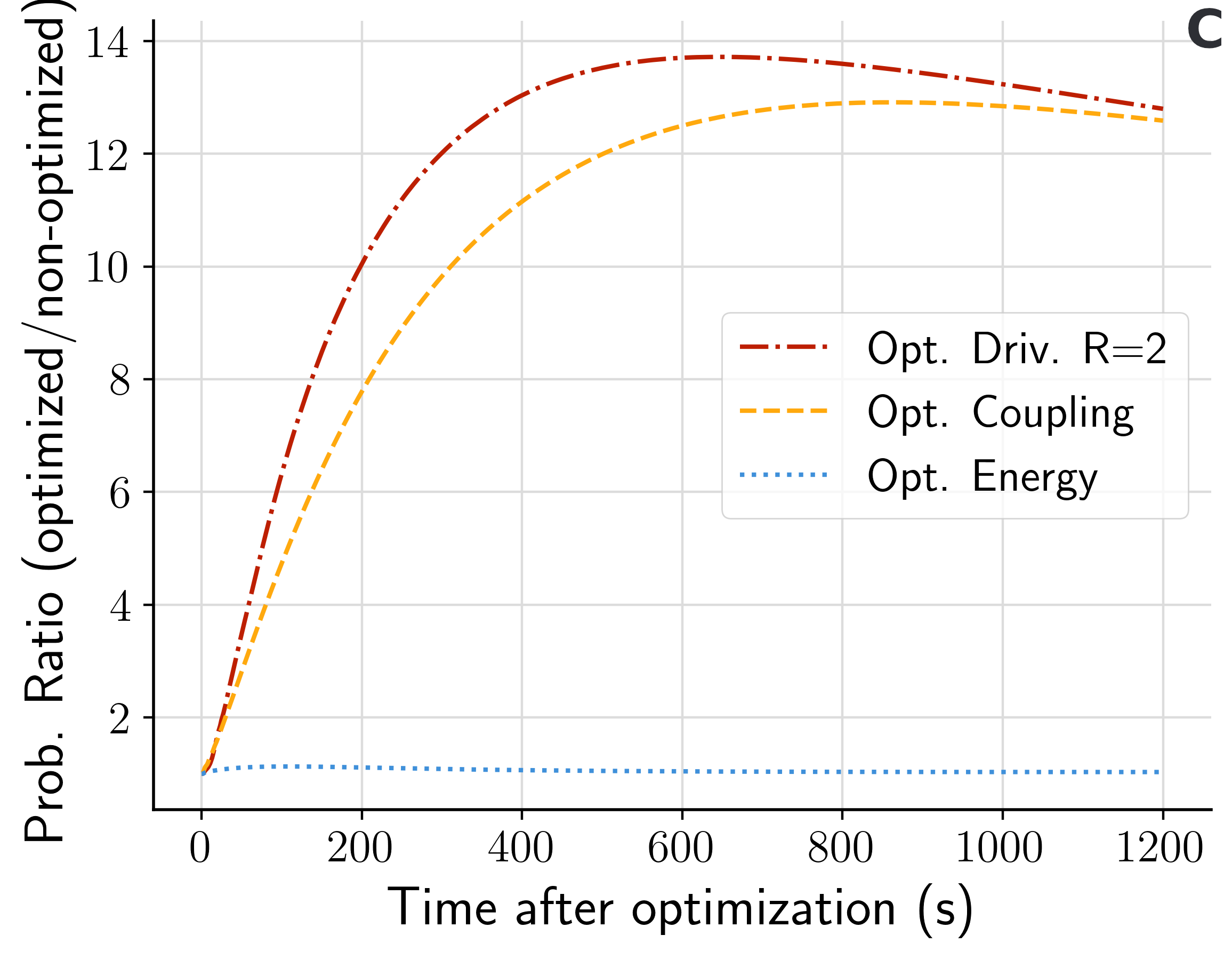}
    \caption{\small Comparison among unoptimised network, optimised driving, optimised coupling, and optimised energies for a SN network of $N=8$ sites in the off-resonant case $\omega_r=15 $. (A) Probability of reaching the sink for the unoptimised network (grey continuous line) and when increasingly expressive drivings ($R=1,2,7$, see Eqs.~\eqref{wq} and \eqref{epsN}) are introduced and optimised (red lines). (B) Probability of reaching the sink for an unoptimised network, optimised driving with $R=2$, optimised coupling, and optimised energies. (C) Ratio between the probability of reaching the sink with the three optimisation strategies and the probability of reaching the sink without optimisation.}
    \label{F04}
    
    \end{figure*}

\autoref{F04}(A) illustrates that, unlike in the NN network, more than one driving term is required for optimal results, i.e. $R > 1$. However, the number of terms remains relatively small, as we observe a significant improvement with just $R = 2$.

In \autoref{F04}(B) and (C) we compare the performance of the unoptimised network against those reached by optimising energies, coupling, and drivings (with $R=2$). The results confirm the conclusions obtained for the NN network: optimising drivings or couplings is a better strategy than optimising the energies of the network.

\subsection{FMO}

In this section we consider our last and more realistic network, a model of the FMO complex \cite{engel2007evidence, cheng2009dynamics}.  

The network Hamiltonian is that found in \cite{sgroi2024efficient}, where the energies are already optimised (see Eq. (A1) of Appendix 1 in \cite{sgroi2024efficient}):
\begin{equation}\label{HFMO}
    H_{N}=\left(\begin{array}{cccccccc}
0 & 0 & 0 & 0 & 0 & 0 & 0 & 0 \\
0 & 65.7 & -104.1 & 5.1 & -4.3 & 4.7 & -15.1 & -7.8 \\
0 & -104.1 & -11.1 & 32.6 & 7.1 & 5.4 & 8.3 & 0.8 \\
0 & 5.1 & 32.6 & -56.1 & -46.8 & 1.0 & -8.1 & 5.1 \\
0 & -4.3 & 7.1 & -46.8 & -36.2 & -70.7 & -14.7 & -61.5 \\
0 & 4.7 & 5.4 & 1.0 & -70.7 & -30.6 & 89.7 & -2.5 \\
0 & -15.1 & 8.3 & -8.1 & -14.7 & 89.7 & 55.7 & 32.7 \\
0 & -7.8 & 0.8 & 5.1 & -61.5 & -2.5 & 32.7 & 4.2
\end{array}\right).
\end{equation}

Most of the conclusions drawn for the previous networks also apply to the FMO. For instance, when we compare the performance of the unoptimised system to cases where we optimised energy, coupling, or driving at the frequency that already maximises the probability of reaching the sink (in this case, $\omega_r = 0.242$), we observe no significant improvements (data not shown).

\begin{figure*}[htbp]
    \includegraphics[width=0.31\textwidth]{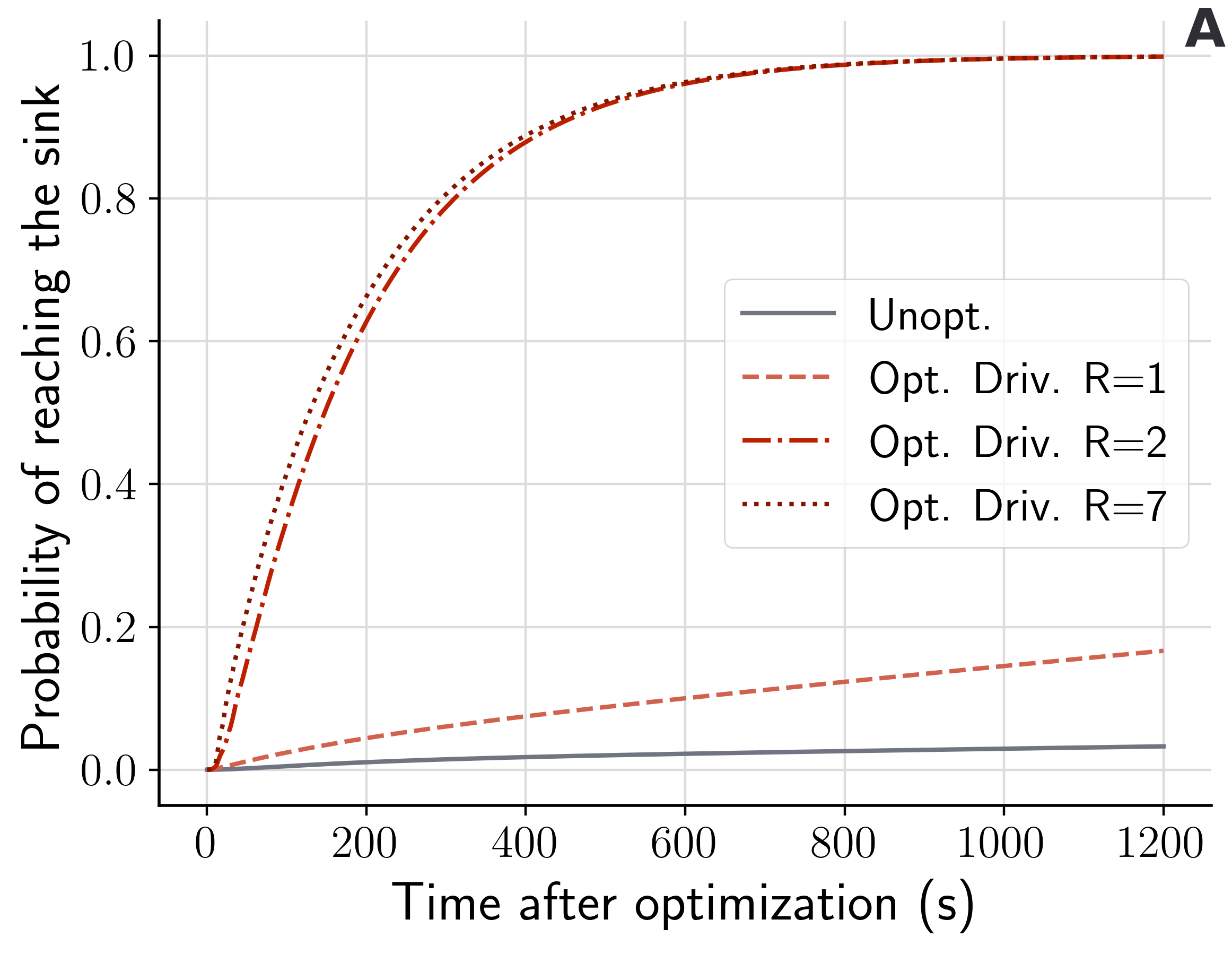}
        \includegraphics[width=0.31\textwidth]{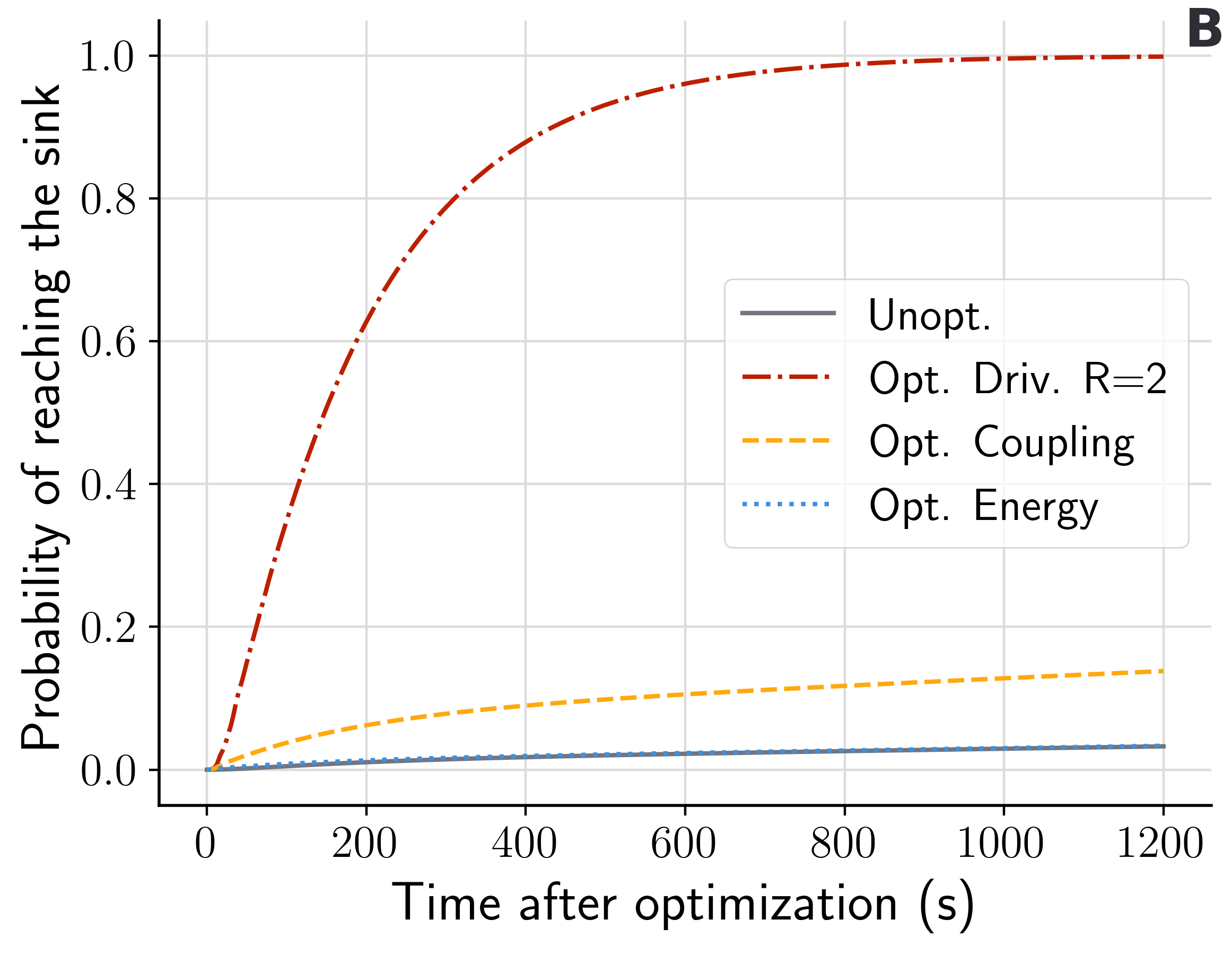}
           \includegraphics[width=0.31\textwidth]{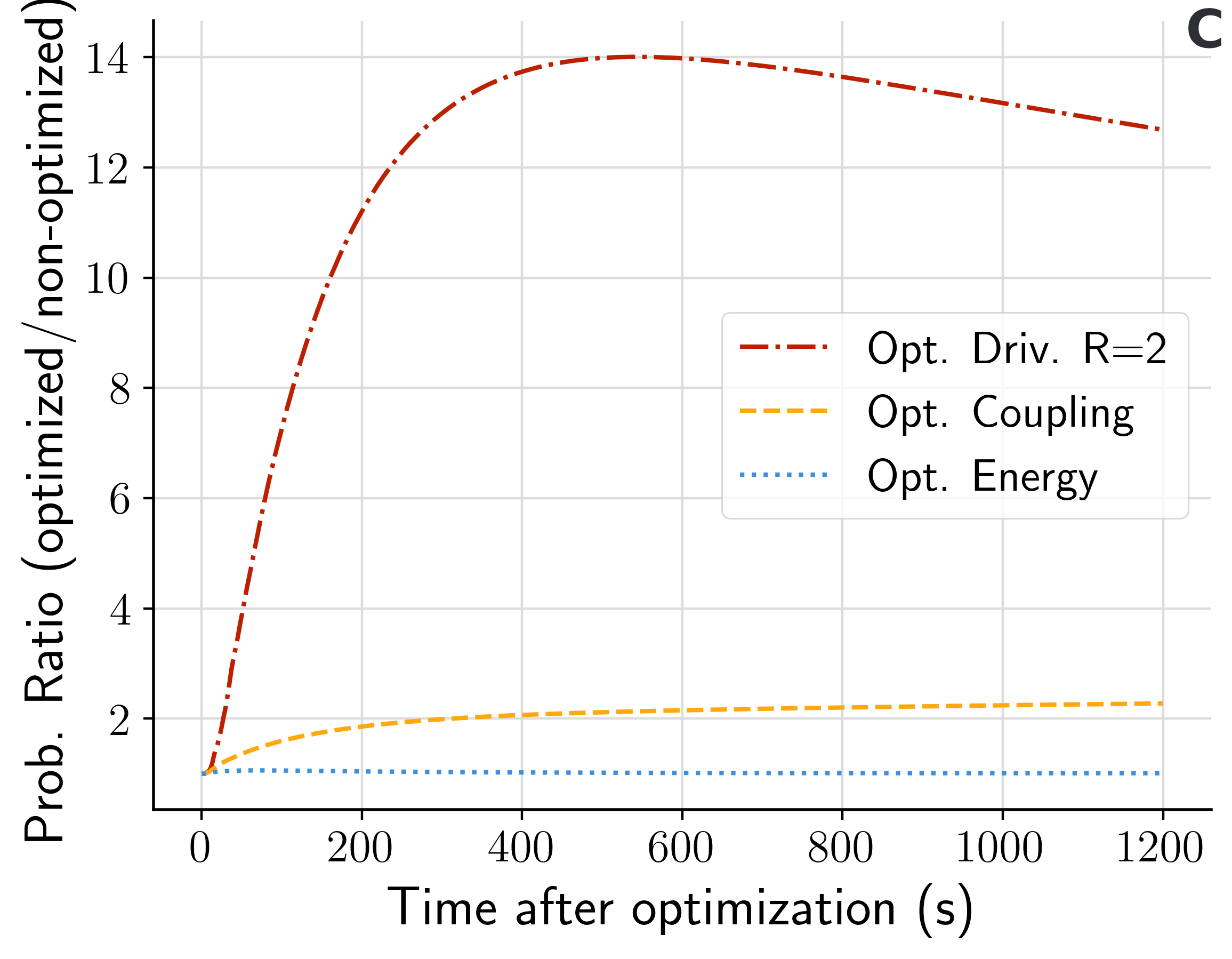}
    \caption{\small Comparison among unoptimised network, optimised driving, optimised coupling, and optimised energies for the FMO network in Eq. \eqref{HFMO} in the off-resonant case $\omega_r=15 $. (A) Probability of reaching the sink for the unoptimised network (grey continuos line) and when increasingly expressive drivings ($R=1,2,7$, see in Eqs.~\eqref{wq} and \eqref{epsN}) are introduced and optimised (red lines). (B) Probability of reaching the sink for an unoptimised network, optimised driving with $R=2$, optimised coupling, and optimised energies. (C) Ratio between the probability of reaching the sink with the three optimisation strategies and the probability of reaching the sink without optimisation.}
    \label{F05}
    
    \end{figure*}

Additionally, \autoref{F05}(A) shows that the driving with one term ($R=1$) is not very effective whereas just adding one term ($R=2$) greatly improves the performance.

In \autoref{F05}(B) and (C) we compare the performance of the unoptimised network against those reached by optimising energies, couplings and drivings (with $R=2$).
The improvement is even more remarkable when one considers that the Hamiltonian of the network $H_N$ in Eq. \eqref{HFMO} was explicitly optimised along the diagonal to maximise exciton transfer in \cite{sgroi2024efficient}.

Also, from the same figures we see that couplings optimisation, differently from the other networks, performs significantly worse than driving optimisation with $R=2$.
This analysis suggests that the optimisation of the drivings can be more effective than that of the couplings, especially when complex networks are considered.

\section{Discussion}

In this work, we investigated the excitation-transfer process across different types of network. In particular we explicitly modelled the radiation-antenna interaction at the absorption stage, and focused on the possible effect of driving on both the antenna and the target site.
Using state-of-the-art automatic differentiation tools and gradient-based optimisation algorithms, we determined the parameters of the system that maximise the efficiency of the transport, as quantified by the time-integrated probability of reaching the sink. 

Our analysis shows that 
optimising the coherent couplings between, respectively, the antenna and the incoming radiation -- and the antenna and the first site of the network --
is effective for simpler networks like NN and SN, but not for more complex networks such as the one associated with the FMO. On the other hand, introducing and optimising external drivings leads to a significant enhancement in excitation transfer across all network types and dimensions. 
Importantly, a small number of driving terms is sufficient to achieve near-optimal efficiency in all considered scenarios: indeed the increase in excitation transfer efficiency in the off-resonant case amounts to more than an order of magnitude compared to the undriven case. This could be relevant in the context of energy harvesting, where a tunable device, capable of efficiently absorbing excitations in a range of several frequencies, is desirable. A deeper understanding of the physical reasons for which learning the couplings is less efficient than introducing and learning optimal drivings 
will be the focus of future research.
Moreover, we expect the driving of the local energies to be more experimentally feasible (e.g. by the modulation of an external field) than acting on the couplings, which might be a more challenging task, also depending on the specific experimental platform one has in mind.

Our analysis concludes that significant enhancement in energy transfer within complex networks can be achieved by applying simple driving modulation at both the start and end of the network. This approach proves especially effective under strong off-resonant conditions between the incoming light and the system, and the practicality of implementing such drivings makes our findings valuable and broadly applicable for optimising energy transfer processes.

\section{Acknowledgements}
We acknowledge financial support from the UK funding agency EPSRC (grant EP/T028424/1), the Royal Society Wolfson Fellowship (RSWF/R3/183013), the Department for the Economy of Northern Ireland under the US-Ireland R\&D Partnership Programme, the ``Italian National Quantum Science and Technology Institute (NQSTI)" (PE0000023) - SPOKE 2 through project ASpEQCt, the “National Centre for HPC, Big Data and Quantum Computing (HPC)” (CN00000013) – SPOKE 10 through project HyQELM, and the EU Horizon Europe EIC Pathfinder project QuCoM (GA no.~10032223). SD acknowledges support from the UKRI through grant EP/X021505/1 and by INFN. AS acknowledges support from MUR under the PON Ricerca e Innovazione 2014–2020 project EEQU and from the PRIN 2022 Project ``Quantum Reservoir Computing (QuReCo)" (2022FEXLYB). 


\section{Appendix 1: Additional case study}\label{appendixA}
In this Appendix, we present findings on additional cases that were not explored in depth in the main text; they all refer to the off-resonant case when $\omega_r=15$.

\begin{figure*}[htbp]
    \includegraphics[width=0.31\textwidth]{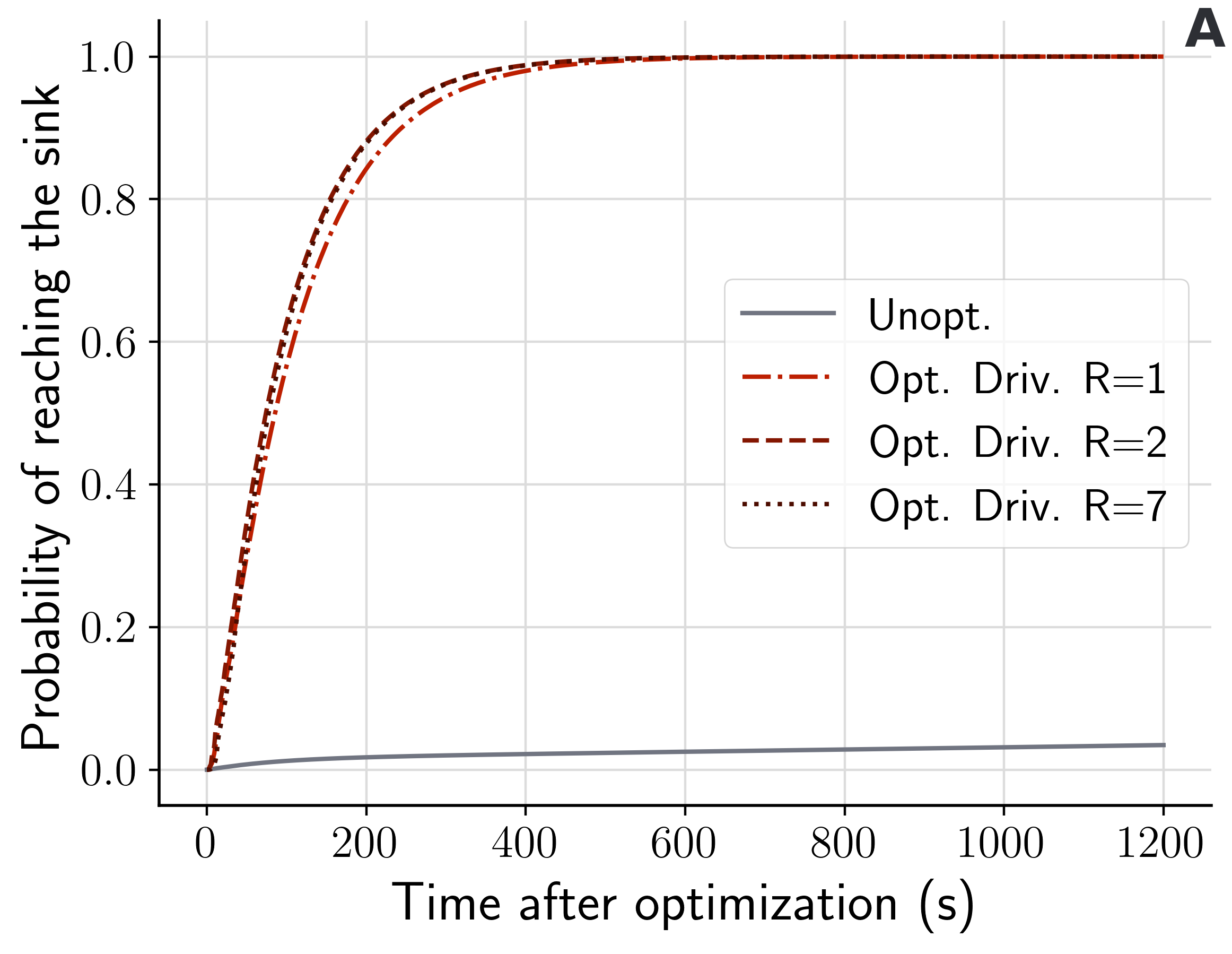}
        \includegraphics[width=0.31\textwidth]{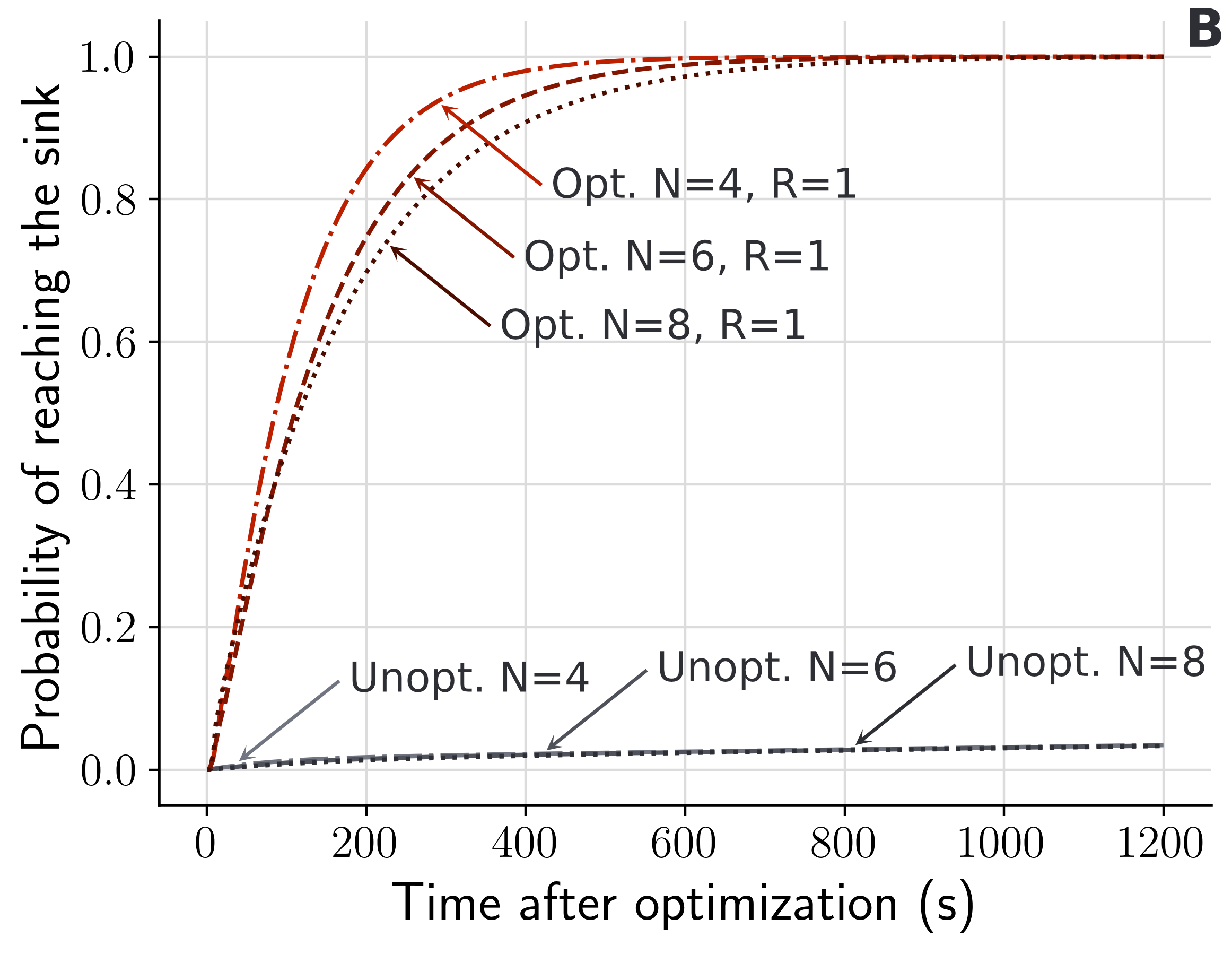}
           \includegraphics[width=0.31\textwidth]{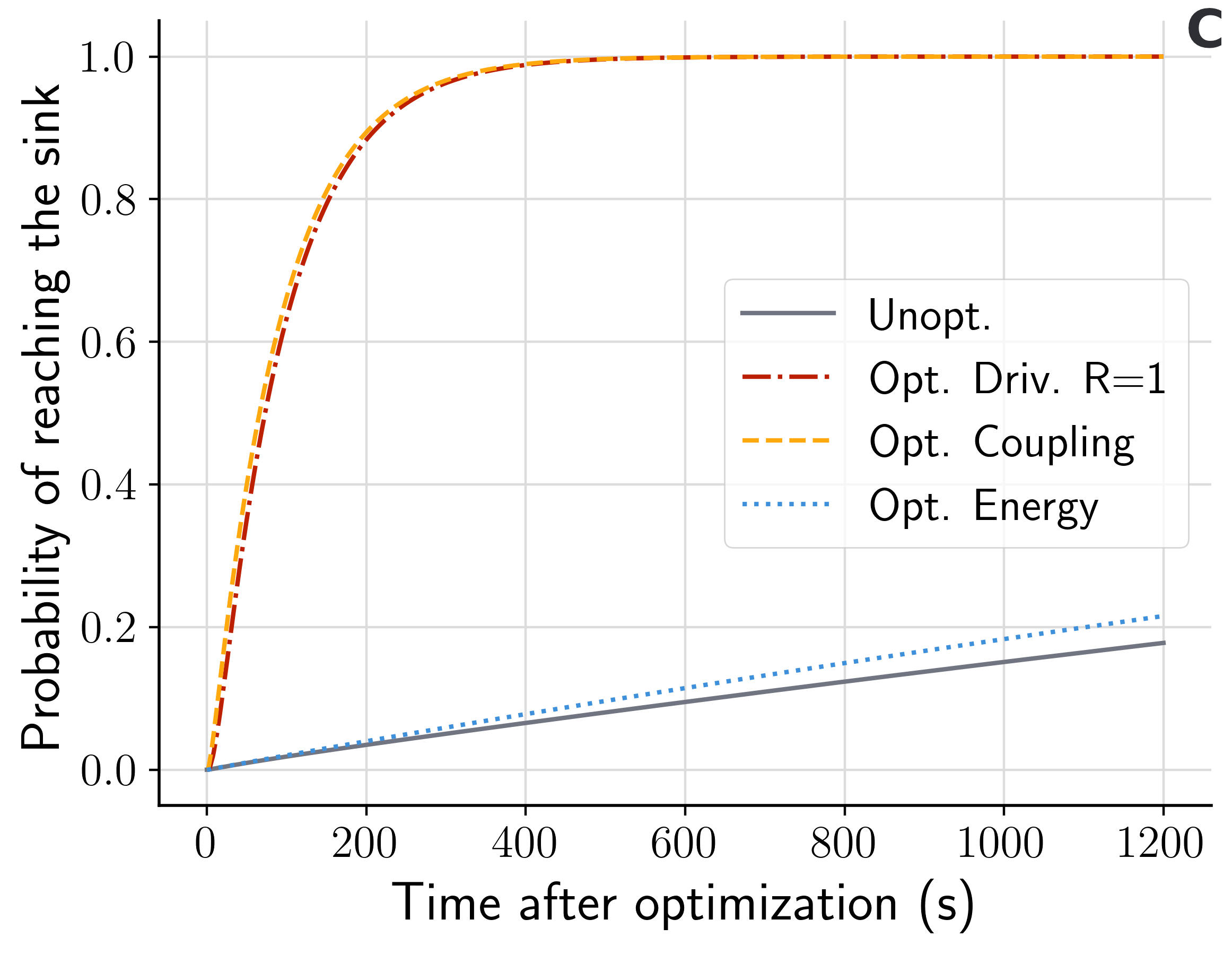}
    \caption{\small For all plots we are in the off-resonant case $\omega_r=15 $. (A) Probability of reaching the sink for a NN network with $N=4$ sites unoptimised (grey continuous line) and when drivings increasingly complex are learnt with $R=1,2,7$ in see Eqs. (\ref{wq}) and (\ref{epsN}). (B) Study of the effectiveness of external driving with $R=1$ for different size of the network $N=4,6,8$. (C) Probability of reaching the sink for a NN network with $N=4$ sites unoptimised, optimising drivings, couplings or site energies with $\lambda_N =1$ one order of magnitude larger than the one considered in \autoref{F02} (A).}
    \label{Fapp}
    
    \end{figure*}

We begin by showing that, when considering the NN network, there is essentially no advantage in using many driving parameters ($R=2,7$) compared to the simplest case ($R=1$), as illustrated in \autoref{Fapp}(A). 

Next, we examine the stability of our results across networks of different sizes. Specifically, we analyse networks with $N=4,6,8$ sites, respectively, and compare the improvement in the probability of reaching the sink when driving is applied. \autoref{Fapp}(B) shows that, although the effectiveness decreases slightly for larger networks, the improvements remain significant in all three cases.

Finally, in \autoref{Fapp}(C) we assess the robustness of our approach when decoherence effects increase by one order of magnitude compared to the case studied in the main text (see \autoref{F02}(A)). The comparison reveals that this increase in decoherence does not affect the conclusions of our analysis; specifically, for the NN network of size $N=4$, learning the couplings or adding driving are the most effective strategies to increase the probability that the excitation reaches the sink. 

\section*{References}

\bibliographystyle{iopart-num}
\bibliography{ref}

\end{document}